\documentclass[final]{svjour2}
\usepackage{graphicx}
\usepackage{rotating}
\usepackage{amssymb}
\usepackage{mathptmx}
\usepackage{amsmath}
\usepackage[numbers]{natbib}
\makeatletter
\journalname{Journal of Low Temperature Physics}

\bibpunct{}{}{,}{s}{}{,}

\begin{document}

\newcommand{\hdblarrow}{H\makebox[0.9ex][l]{$\downdownarrows$}-}
\title{Phonon Pulse Shape Discrimination in SuperCDMS Soudan}

\author{S.A. Hertel$^1$ and M. Pyle$^2$ for the SuperCDMS Collaboration}

\institute{1: Massachusetts Institute of Technology, Department of Physics, Cambridge, MA 02189, USA
\email{hertel@mit.edu}
\\2: Stanford University, Department of Physics, Stanford, CA 94305, USA}

\date{07.08.2011}

\maketitle


\begin{abstract}
SuperCDMS is the next phase of the Cryogenic Dark Matter Search experiment, which measures both phonon and charge signals generated by particle recoils within a germanium target mass.  Charge signals are employed both in the definition of a fiducial volume and in the rejection of electron recoil background events.  Alternatively, phonons generated by the charge carriers can also be used for the same two goals.  This paper describes preliminary efforts to observe and quantify these contributions to the phonon signal and then use them to reject background events.  A simple analysis using only one pulse shape parameter shows bulk electron recoil vs. bulk nuclear recoil discrimination to the level of 1:10$^3$ (limited by the statistics of the data), with little degradation in discrimination ability down to at least 7~keV recoil energy.  Such phonon-only discrimination can provide a useful cross-check to the standard discrimination methods, and it also points towards the potential of a device optimized for a phonon-only measurement.

\end{abstract}

\section{Phonons in a SuperCDMS Detector}

SuperCDMS (Cryogenic Dark Matter Search) Soudan is in preparation for installation at the Soudan Underground Laboratory in northern Minnesota.  The target mass consists of $\sim$15 large roughly cylindrical (3.81~cm radius, 2.54~cm thickness, $\sim$600~g) ultra-pure germanium crystals.  These target masses are instrumented on the flat surfaces with interleaved arrays of phonon-sensing Quasiparticle-trap-assisted Electrothermal-feedback Transition-edge-sensors (QETs) and charge-sensing electrodes.\cite{luke1994, brink2006}  The measurement of both the phonon energy and charge energy of each event enables the ratio of these two energies to be used as a powerful event-by-event discriminator between low-charge-production nuclear recoils (a possible dark matter signature) and high-charge-production electron recoils (the vast majority of background events).


The interleaving of biased electrodes (typically +2~V on one side, -2~V on the opposite side) with QET arrays (essentially at 0~V) defines a $\sim$0.5~V/cm drift field in the bulk of the target mass.  Within $\sim$1~mm of the top and bottom surfaces, however, the field created by the interleaved structures is both much stronger than the bulk ($\sim$20~V/cm), and also largely parallel to the surface.  This field shape prevents carriers created in this region from propagating to the opposite side electrodes, and the resulting side-asymmetric charge signal tags such events as near-surface events\cite{pyle2009} (near-surface electron recoils would otherwise form a dangerous background\cite{bailey2009}).

Much of the essential information content of the charge measurement is duplicated in the phonon measurement, through the production of secondary phonon populations as charges propagate through the crystal and then enter aluminum structures at the surface.  These different phonon populations are produced with different initial energy distributions.  In Germanium, high energy phonons exhibit high isotopic scattering rates ($\tau_{i}^{-1}~=~[36.7\times10^{-42}]\nu^4$ for Ge)\cite{tamura1985} meaning that they have short mean free paths and propagate diffusively ($\ell~=~[5.4~km/s]\tau_{i}~=~[1.5\times10^{44}~m/s^4]\nu^{-4}$ at long wavelengths).  Only 6.1$\%$ of the top and bottom crystal surfaces is covered with phonon-absorbing aluminum (largely in the form of QETs), meaning that, per surface interaction, a phonon is far more likely to reflect off a polished Ge surface than be absorbed in a QET.  A phonon's timescale of surface absorption then is roughly proportional to its rate of surface interactions, and is therefor highly dependent on mean free path.  Phonons with a mean free path on order of the detector size (ie, which propagate ballistically) are absorbed with the maximally slow decay constant of 755~$\mu$s, and higher energy phonons are absorbed with significantly faster decay constants.  Phonons anharmonically decay in energy ($\tau_{a}^{-1} = [1.61\times10^{-55}]\nu^5$)\cite{msall1997}, meaning that mean free path increases with time.  This energy-dependent diffusion and decay process is typically termed ``quasidiffusion''.  Here we summarize the relevant characteristics of the major phonon populations in a SuperCDMS detector, emphasizing where they begin in their quasidiffusive evolution:

\begin{itemize}
\item \textbf{Primary phonons} produced by the recoil event itself are initially highly energetic ($\nu > $1~THz, ie $\ell~<~\sim$~100$\mu$m).  If the event occurs far from a sensor surface, the diffusive behavior slows the arrival at the surface and lengthens the eventual mean free path, slowing the absorption rate at that surface.  If the event occurs near a sensor surface, there is little delay in arrival, and the comparatively short mean free path increases a the rate of absorption at that surface.
\item \textbf{Neganov-Luke phonons}\cite{neganov1978, luke1988} are created as charge carriers are drifted by the electric field.  Such phonons are created very soon after the event time (the charge drift time is $\sim$~1~$\mu$s).  In the low field of the detector bulk, Neganov-Luke phonons are emitted with low energies and correspondingly ballistic free paths, whereas the majority of Neganov-Luke phonons are produced in the strong near-surface fields at higher frequencies ($\sim$300 GHz and $\sim$700 GHz for electrons and holes) with correspondingly short mean free paths ($\sim$1.8~cm and $\sim$600~$\mu$m)\cite{wang2010, leman2010} and fast absorption rates.
\item \textbf{Relaxation phonons} are emitted when charge carriers cross from the Ge into the superconducting Al surface structures, converting their Ge gap energy (0.75 eV) into Al quasiparticles (of energy 2$\Delta_{Al}$~=~200~$\mu$eV)\cite{court2008} and a population of initially high energy phonons.  In the presence of superconducting Al, such phonons experience rapid downconversion (at time scales of $\sim$~1~$\mu$s)\cite{kaplan1976, kozorezov2007}, meaning that relaxation phonons are ballistic at the time scales considered here.
\end{itemize}

These three phonon populations combine together to produce the observed phonon pulses and their associated rising and falling timing characteristics.  We will from now on ignore relaxation phonons, since they are initially ballistic (their absorption time constant is a slow $\sim$755~$\mu$s, and a minimal contribution to pulse shape).  A recoil occurring in the crystal bulk will produce a roughly side-symmetric signal of near-surface Neganov-Luke phonons, which will arrive at both the top and bottom surfaces at fast times ($\sim$1~$<~t~<~\sim$5~$\mu$s) and are absorbed rapidly by these surfaces.  Primary phonons will reach the top and bottom surfaces more gradually, as quasi-diffusive propagation allows ( $\sim$1~$<$~t~$< \sim$15~$\mu$s).  These primary phonons will have shorter mean free paths (and will be more quickly absorbed) at the closer surface, and will have longer mean free paths (and will be more slowly absorbed) at the further surface.  When an event occurs within the near-surface electric field region, it will be distinguished by a boosted population of Neganov-Luke phonons at one surface (with fast absorption times) and a lack of any distinct Neganov-Luke sharp rising edge in the opposite side pulse. 


\newlength{\oldbelowdisplayskip}
\newlength{\oldabovedisplayskip}
\setlength{\oldbelowdisplayskip}{\belowdisplayskip}
\setlength{\oldabovedisplayskip}{\abovedisplayskip}
\setlength{\belowdisplayskip}{0.5em}
\setlength{\abovedisplayskip}{0.5em}
\setlength{\belowdisplayskip}{\oldbelowdisplayskip}
\setlength{\abovedisplayskip}{\oldabovedisplayskip}

\section{Quantifying Phonon Pulse Characteristics in SuperCDMS Detectors}

The understanding of phonon populations given in the previous section predicts certain shape characteristics as a function of both proximity to the absorbing surface and quantity of charge carriers (and resulting Neganov-Luke production).  Looking at recent calibration data (Figure 1), we do see such behaviors, confirming our understanding.
\begin{figure}
\begin{center}
\includegraphics[scale=0.13]{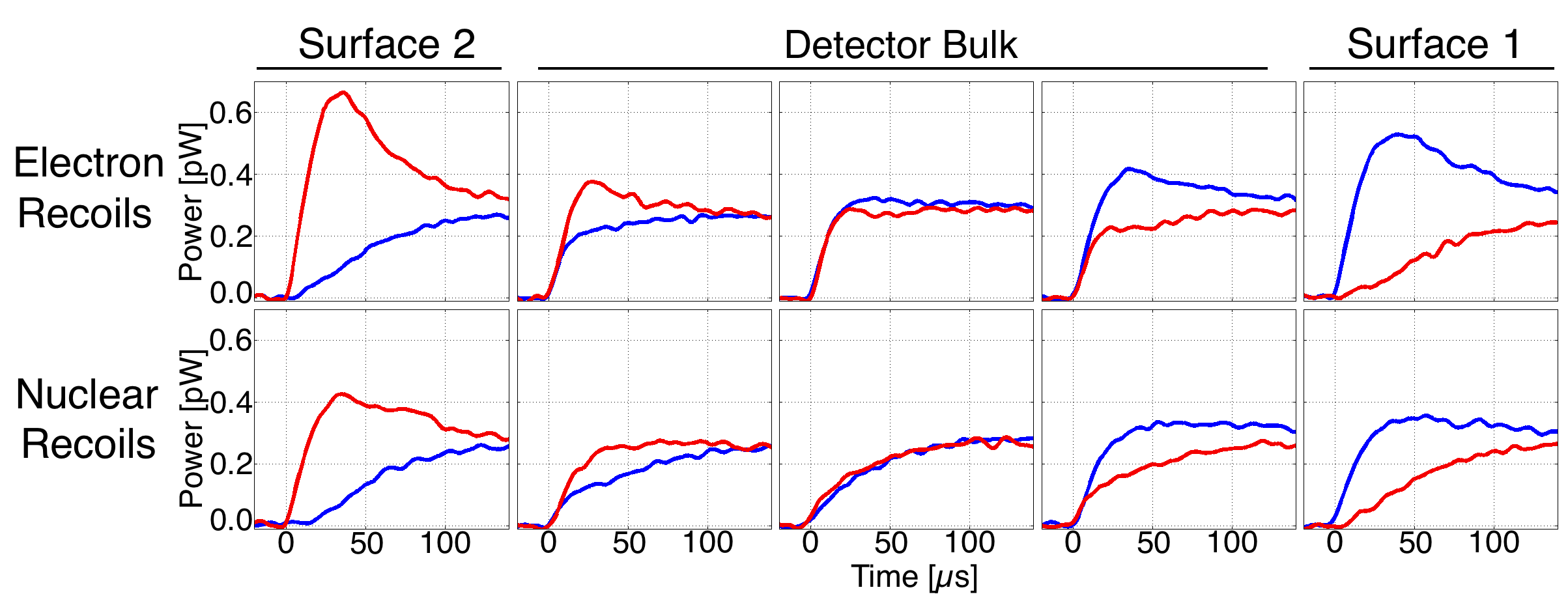}
\caption{Example phonon pulses from calibration runs with a SuperCDMS Soudan detector.  The four side~1 QET channels are summed and shown in blue (dark grey); the side~2 channels are summed and shown in red (light grey).  All eight events shown are of similar total phonon energy ($\sim$50 keV), and are located at different positions along the central axis of the cylindrical crystal (as determined by relative weighting among the eight phonon channels).  Surface~1 and surface~2 events were selected using an additional a requirement that the charge signals from the two sides be strongly asymmetric.\label{fig:pulses}}
\end{center}
\end{figure}

To make use of these phonon pulse shape variations as event type discriminators(distinguishing electron recoil and near-surface recoils from nuclear recoils), we first reduce the raw traces to a small number of simple quantities, and then combine these quantities in a way that accommodates the large position-dependent variation.  First, each of the raw traces from the detector's eight QET channels were fit using the following functional form:
\setlength{\oldbelowdisplayskip}{\belowdisplayskip}
\setlength{\oldabovedisplayskip}{\abovedisplayskip}
\setlength{\belowdisplayskip}{0.5em}
\setlength{\abovedisplayskip}{0.5em}
\begin{align}
P(t) = A_{fast} \bigg(1-e^{\frac{-(t-T_{fast})}{R_{fast} }}  \bigg) e^{\frac{-(t-T_{fast})}{F_{fast} }}  +
          A_{slow}\bigg(1-e^{\frac{-(t-T_{slow})}{R_{slow} }}  \bigg) e^{\frac{-(t-T_{slow})}{F_{slow} }} \nonumber
\end{align}
\setlength{\belowdisplayskip}{\oldbelowdisplayskip}
\setlength{\abovedisplayskip}{\oldabovedisplayskip}
where the total pulse is treated as the sum of a ``fast'' pulse and a ``slow'' pulse, with rising and falling time constants $R_{fast}$, $R_{slow}$, $F_{fast}$, and $F_{slow}$, and start time offsets $T_{fast}$ and $T_{slow}$.  The fast pulse can be thought of as the contribution of highly-diffusive quickly-absorbing phonons, while the slow signal can be thought of as the contribution of more ballistic slowly-absorbing phonons, but the quasi-diffusive nature means that in reality there is not a simple two-category distinction. The falling time constant $F_{slow}$ of the slow pulse was observed to be nearly identical for every event, and was set to 755~$\mu$s, the observed rate of absorption of the late-time uniform bath of low energy phonons.\cite{kevin2011}  Each event's resulting 7 fit parameters for each of 8 phonon channels were further reduced by summing the fits for each side and then finding key points (10$\%$,  20$\%$,  30$\%$,  etc.) along the rising and falling edges of these side-summed fits.

The partition of energy between the eight channels also contains significant discrimination information (in addition to position information).  The amplitude of each channel's pulse was obtained using an optimal filter, in which a template pulse of fixed shape was scaled to best match the amplitude of the (variable shape) phonon pulse.  These measured amplitudes, then, are shape-dependent.  Distributions of some example partition and pulse shape quantities are shown in Figure 2.
\begin{figure}
\begin{center}
\includegraphics[scale=0.15]{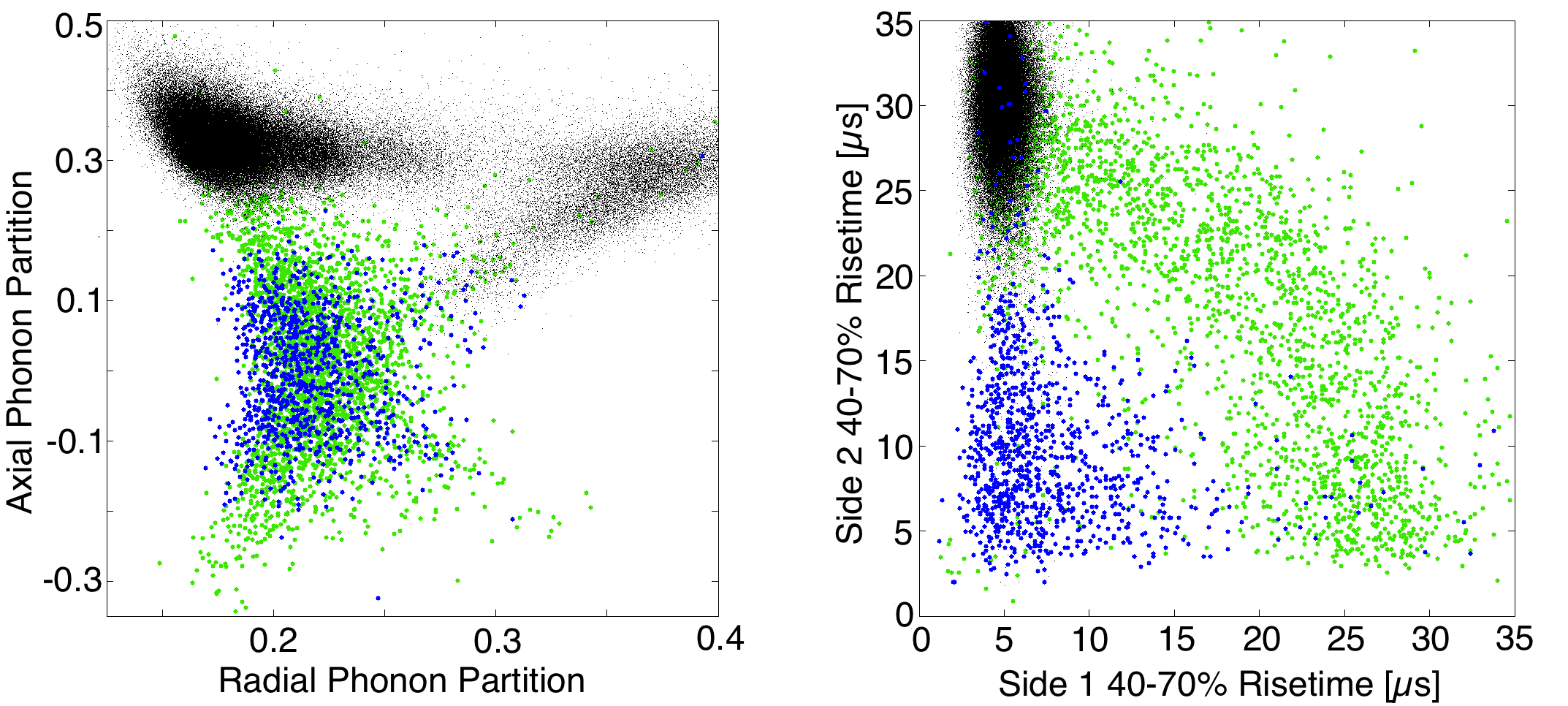}
\caption{Example phonon discrimination quantities.  On the left, an example of partitioning between channels is shown.  The vertical axis is an axial partition quantity (side1 - side2)/(side1 + side2), and the x~axis is a similarly defined radial partition quantity.  On the right, the time difference (in $\mu$s) between the 40$\%$ and 70$\%$ points on the rising edge of the side~1 summed pulse (x~axis) and side~2 summed pulse (y~axis) are shown.  In both plots, calibration events are colored using a charge-signal-based categorization as either bulk electron recoils (blue or dark grey), bulk nuclear recoils (green or light grey), or side~1 electron recoils (black).  The total phonon energy for these events are between 7 and 20~keV recoil energy, where recoil energy has been scaled from total phonon energy using a nuclear recoil assumption for all events.\label{fig:rqdistributions}}
\end{center}
\end{figure}

\section{Discrimination Based on Phonon Pulse Quantities}

As a very first look at discrimination based on phonon pulse shape characteristics, we here show a simple example of bulk electron recoil (ER) vs bulk nuclear recoil (NR) discrimination using only one pulse shape characteristic: the 40$\%$-to-70$\%$ risetimes.  This simple discrimination example serves as a lower bound on the abilities one might expect when using a combined analysis of all phonon pulse quantities.

We first construct ER and NR populations.  Identical charge-based fiducial volume cuts are enforced on both populations, and measured [charge:phonon] energy ratios are used to categorize events as either ER or NR.  After these two populations have been defined, a new pulse shape discrimination quantity is constructed using the 40$\%$-to-70$\%$ risetimes (for side~1 and side~2) for each event:  ``radius'' = ([40$\%$-to-70$\%$ side 1]$^2$ + [40$\%$-to-70$\%$ side 2]$^2$)$^{1/2}$.  One can see in Figure 2 that this combined quantity is largely position-independent.

Figure 3 shows a histogram of this timing quantity, plotting only the lowest energies inspected in this analysis (7~$<$~E$_{recoil}$~$<$~20~keV, defined by assuming that all events are nuclear recoils and scaling the total phonon energy accordingly).  Although the ER and NR distributions overlap somewhat, there are no slow ER outliers to the statistics available in the calibration dataset.  The right panel of Figure 3 shows the NR acceptance fraction vs ER leakage fraction as one varies a 1D cut threshold.  Discrimination better than 1:10$^3$ is seen, and it is further seen that this discrimination shows no degradation with energy down to at least 7~keV~E$_{recoil}$.

\begin{figure}
\begin{center}
\includegraphics[scale=0.25]{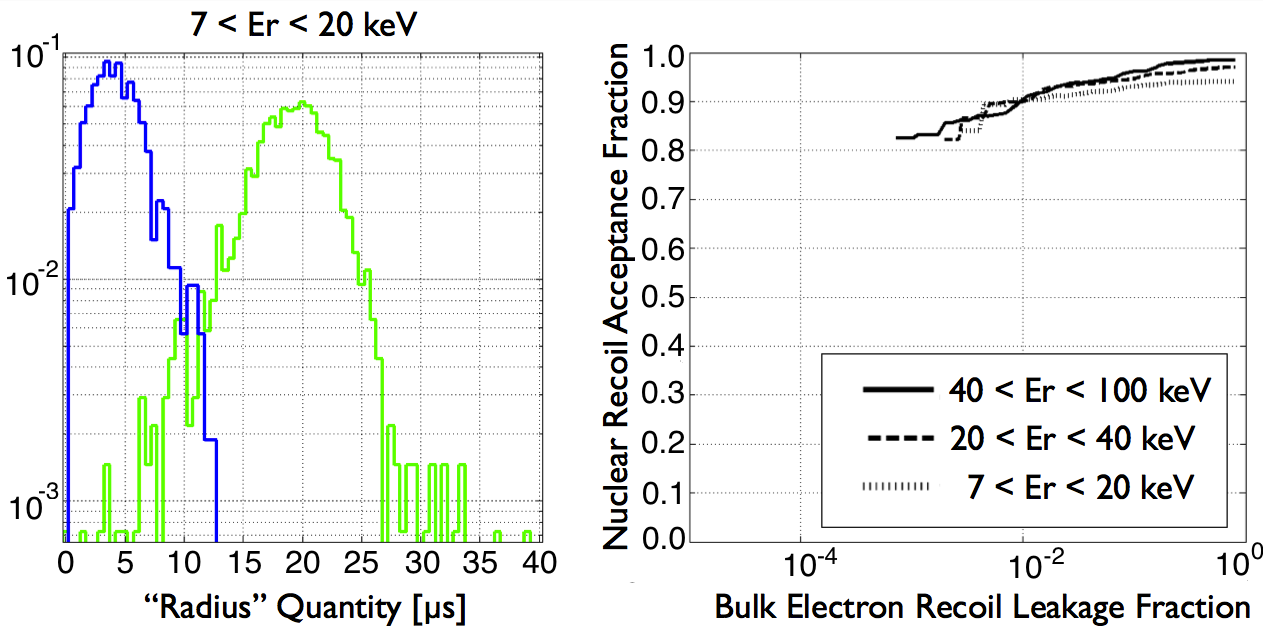}
\caption{A simple look at pulse shape discrimination between bulk nuclear recoils and bulk electron recoils, using 40$\%$-to-70$\%$ risetimes.  Nuclear recoils are green (light grey); electron recoils are blue (dark grey).  For discussion, see text.\label{fig:discrimination}}
\end{center}
\end{figure}

\section{Conclusions}

Although significant pulse shape discrimination ability has been demonstrated, phonon-only discrimination is only beginning to show its promise.  The simple analysis performed here can be improved in many ways, and looking further on the horizon, this rough analysis using a detector optimized for charge-electrode-based discrimination suggests that a detector optimized specifically for phonon-only discrimination could take advantage of the extreme sensitivity of transition edge sensors to extend event-by-event discrimination well below the typical CDMS low energy threshold of $\sim$10~keV (set by charge readout noise).  Such phonon-only discrimination capabilities should be possible if detectors are specifically optimized for this goal, through the reduction of TES internal thermal fluctuation noise\cite{adam2011}, and also through an increase in pulse shape differences themselves by increasing the total phonon-absorbing Al area of the QETs.


\begin{acknowledgements}
We would like to thank E. Figueroa-Feliciano, S.W. Leman, and Adam Anderson for valuable discussions.  This work is supported by the United State Department of Energy under grand DE-AC02-76SF00515 and by the United States National Science Foundation under grants PHY-0847342, 0705052, 0902182, and 1004714.
\end{acknowledgements}


\end{document}